\documentclass[%
 reprint,
 superscriptaddress,
 amsmath,amssymb,
 aps,
 prl,
]{revtex4-1}

\usepackage{graphicx}
\usepackage{dcolumn}
\usepackage{bm}
\usepackage{multirow}
\usepackage{times}

\newcommand{\re}{\rm{Re}}
\newcommand{\fb}{\bar{f}}
\newcommand{\hb}{\bar{h}}
\newcommand{\dl}{\mathcal{L}}

\begin{document}

\preprint{APS/123-QED}

\title{A Bayesian machine scientist to compare data collapses for the Nikuradse dataset}

\author{Ignasi Reichardt}
 \email{ignasi.reichardt@urv.cat}%
\affiliation{%
  Department of Chemical Engineering, Universitat Rovira i Virgili, Tarragona 43007, Catalonia, Spain}%
\author{Jordi Pallar\`es}%
 \email{jordi.pallares@urv.cat}%
 \affiliation{%
  Department of Mechanical Engineering, Universitat Rovira i Virgili, Tarragona 43007, Catalonia, Spain}%
\author{Marta Sales-Pardo}%
 \email{marta.sales@urv.cat}%
\affiliation{%
  Department of Chemical Engineering, Universitat Rovira i Virgili, Tarragona 43007, Catalonia, Spain}%
 \author{Roger Guimer\`a}%
  \email{roger.guimera@urv.cat}%
 \affiliation{%
    Department of Chemical Engineering, Universitat Rovira i Virgili, Tarragona 43007, Catalonia, Spain}%
 \affiliation{ICREA, Barcelona 08010, Catalonia, Spain}

\date{\today}

\begin{abstract}
Ever since Nikuradse's experiments on turbulent friction in 1933, there have been theoretical attempts to describe his measurements by collapsing the data into single-variable functions. However, this approach, which is common in other areas of physics and in other fields, is limited by the lack of rigorous quantitative methods to compare alternative data collapses. Here, we address this limitation by using an unsupervised method to find analytic functions that optimally describe each of the data collapses for the Nikuradse data set. By descaling these analytic functions, we show that a low dispersion of the scaled data does not guarantee that a data collapse is a good description of the original data. In fact, we find that, out of all the proposed data collapses, the original one proposed by Prandtl and Nikuradse over 80 years ago provides the best description of the data so far, and that it also agrees well with recent experimental data, provided that some model parameters are allowed to vary across experiments. 

\end{abstract}

\maketitle

\begin{figure}
  \includegraphics[width=\columnwidth]{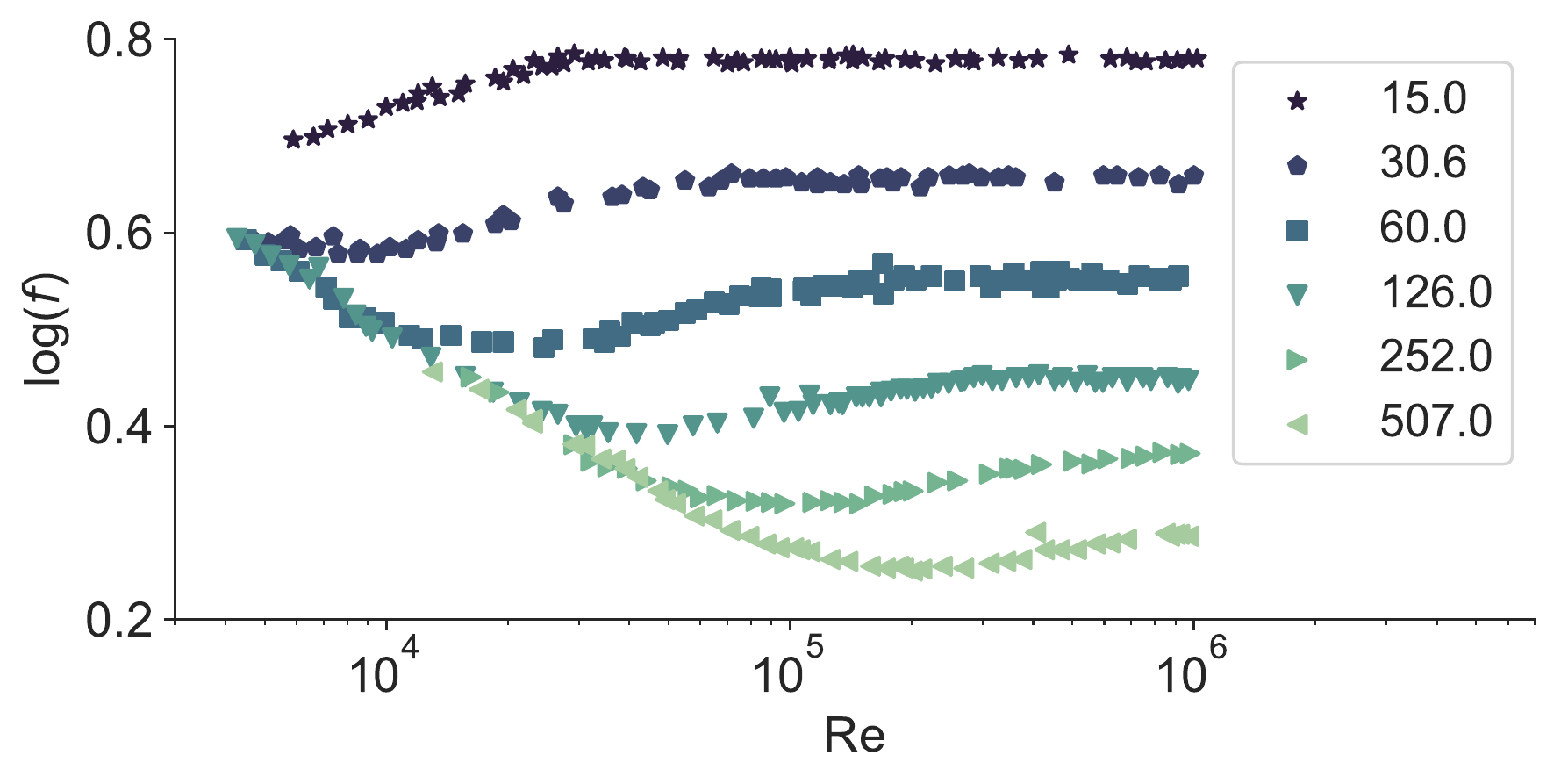}
  \vspace{-8mm}
  \caption{The Nikuradse dataset, displaying the turbulent friction factor $f$ as a function of the Reynolds number $\re$ for different values of the inverse relative roughness $k/r$, which are indicated in the legend.}
\label{fig:niku}
\end{figure}

In the early 1930s, Johann Nikuradse conducted experiments to measure the friction sustained by a turbulent flow in a rough pipe \cite{nikuradse1933laws}. With remarkable accuracy, he measured the dependency of the friction factor $f$ on two dimensionless quantities, the Reynolds number $\re$ and the relative roughness, that is, the ratio $r/k$ between the size of the irregularities and the radius of the pipe (Fig.~\ref{fig:niku}). Over eight decades later, and despite the fundamental and practical importance of the problem, the functional relationship $f=h(\re , r/k)$ remains unknown.

Numerous works \cite{nikuradse1933laws,prandtl,Note1,goldenfeld2006roughness,tao2009critical,she2012multi,li2016united} have attempted to solve the problem by collapsing Nikuradse's data into a function $\fb = \hb(x)$ that depends on a single variable $x$ combining both the Reynolds number (or the so-called turbulent Reynolds number, $\re_{\tau}$) and the relative roughness. For example, Prandtl and Nikuradse~\footnote{In his original paper, Nikuradse already presented his data scaled according to Prandtl's collapse. However, this collapse is mentioned in~\cite{she2012multi}, where Bodenschatz attributes it to Prandtl, Nikuradse's supervisor.} proposed the collapse $f^{-1/2} + 2 \log(r/k) = \hb(\re_{\tau}\cdot r/k)$, which suggests that a transformed turbulent friction factor only depends on the product of the turbulent Reynolds number and the relative roughness \cite{nikuradse1933laws,prandtl}.

Data collapses such as this are common in many areas, and aim at establishing the combinations of independent variables that actually affect the dependent variable, disregarding the exact functional form of the dependency. These approaches have been particularly useful in critical phenomena, because the scaling behavior of critical systems implies that, close to the critical point, all relevant functions must be generalized homogeneous functions and therefore collapse under particularly simple scaling transformations \cite{stanley99,barenblatt03}. This realization has led to important insights in physics and other disciplines, even for systems that are not at a critical point. For example, data collapses have enabled the characterization of the statistical laws that govern the growth of human organizations \cite{stanley96,lee98}.

\begin{figure*}
\centering
\includegraphics[width=\textwidth]{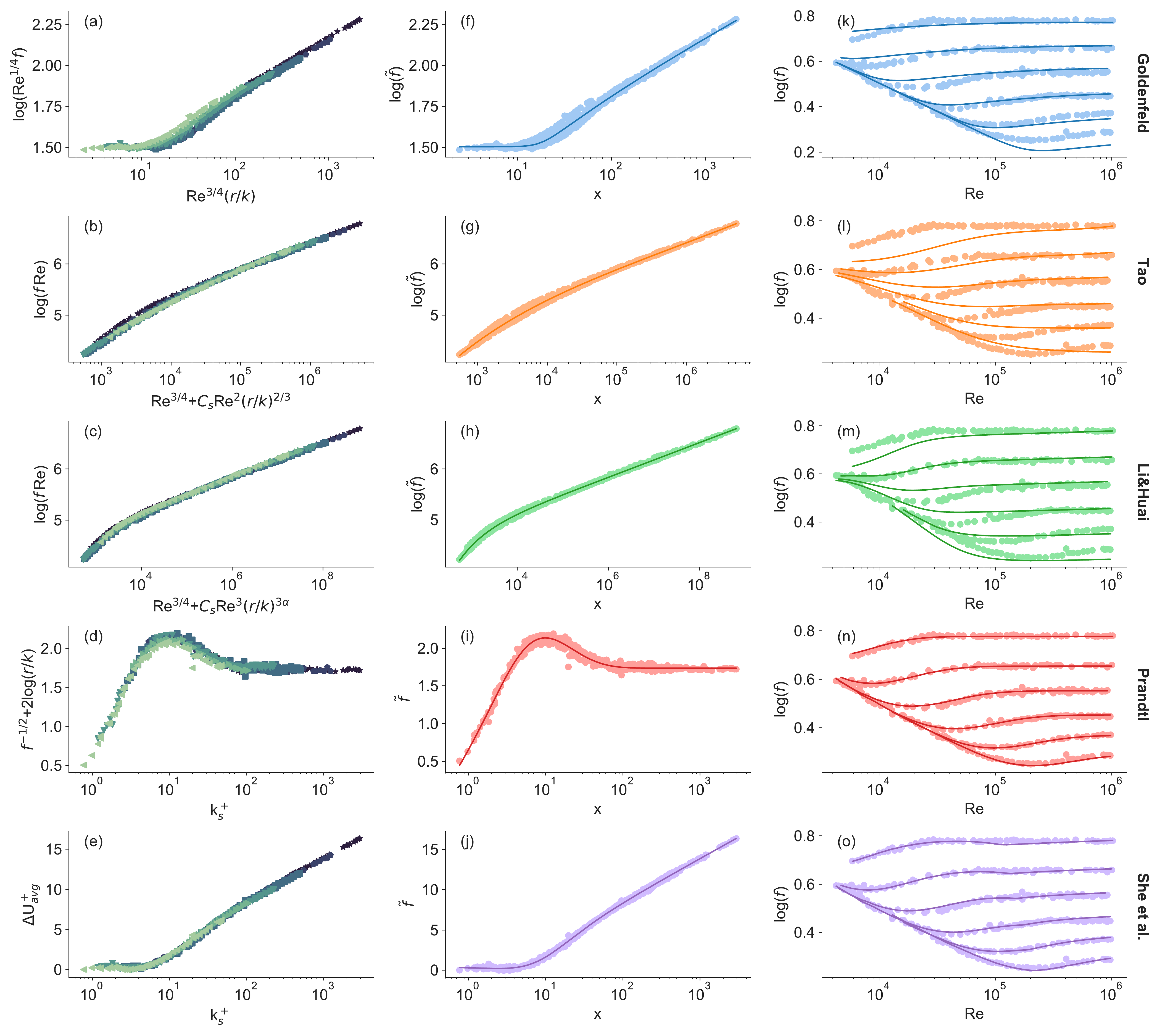}
\vspace{-8mm}
\caption{(a)-(e) Nikuradse data collapsed according to each approach (Goldenfeld \cite{goldenfeld2006roughness}, Tao \cite{tao2009critical}, Li and Huai \cite{li2016united}, Prandtl \cite{prandtl}, and She et al. \cite{she2012multi}).
(f)-(j) Models for the collapsed data. The solid line represents the most plausible closed-form mathematical model identified by the Bayesian machine scientist for each collapse.
(k)-(o) Unscaled models. Same as the middle column but unscaled to the original variables.}
\label{fig:scalings}
\end{figure*}

Perhaps the major challenge of approaches based on scaling and data collapse is the fact that, in principle, many alternative collapses are possible, especially when the data are noisy. In the Nikuradse dataset, for example, different arguments may lead to very different collapses (Fig.~\ref{fig:scalings}, Tab.~\ref{tab:scalings}). In such cases, the goodness of a collapse is typically evaluated qualitatively by how well the empirical data obtained under different conditions fall onto a single well-defined ``scaling function'' $\hb$ \footnote{We refer to $\hb$ as the {\em scaling function} although, strictly speaking, this function is not always a generalized homogeneous function.}. In the few remarkable instances in which the goodness of the collapse is quantified, existing methods rely on interpolation of the datasets to approximate the scaling function, and on measuring deviations in the collapsed data \cite{bhattacharjee01}. Additionally, they allow only to compare a scaling function with different parameter values, but not to compare scaling functions that are mathematically different.

Here, we argue that the goodness of a data collapse should be measured on the original unscaled data because: (i) the scaling function distorts the data by stretching some regions and compressing others, thus potentially obscuring deviations in the collapse; (ii) the distortion can lead to collapse functions with very different ranges, thus making comparison between collapses meaningless. 
We propose a rigorous alternative to evaluate the quality of data collapses, and show that scaling functions that seem better at collapsing the data often do not describe the original unscaled data well.
%
%
Our approach proceeds by first obtaining the most plausible functional form for the scaling function $\hb$, for which we use an unsupervised algorithm that explores systematically the space of possible functional forms for $\hb$; following Refs.~\cite{evans10,robot} we call this algorithm a machine scientist. Next, we unscale this function and quantify how it fits the original data, as opposed to the collapsed data. For the Nikuradse dataset, we find that the best collapse is the one originally proposed by Nikuradse and Prandtl \cite{nikuradse1933laws,prandtl}. We also study what is the applicability of the best models to recent datasets on turbulent friction in rough pipes. 

\medskip


\begin{table*}[t!]
  \caption{Summary of the data collapses for the Nikuradse dataset. }
  \begin{ruledtabular}
\begin{tabular}{cccc}
\label{tab:scalings}
Reference & Scaling variable, $x$ & Scaling function, $\hb$ & Most plausible model, $h(x)$ \\
\hline
Goldenfeld~\cite{goldenfeld2006roughness} & Re$^{3/4}r/k$ & $f$Re$^{1/4}$ & $\frac{1}{c_{3}} \log{\left (\left(c_{1} x\right)^{c_{2} + x} + e^{c_{1}} \right )}$ \\
Tao$^a$~\cite{tao2009critical}        & Re$^{3/4}+C_s$Re$^2(r/k)^{2/3}$    & $f$Re & $c_{1} \left(c_{2} c_{2}^{- x} x + c_{3}\right) + x$ \\
Li\&Huai$^b$~\cite{li2016united}      & Re$^{3/4}+C_s$Re$^3(r/k)^{3\alpha}$ & $f$Re & $c_{1} \left(c_{2} + c_{3} x \left(c_{1} + c_{4}^{x}\right)\right)$ \\
Prandtl$^c$~\cite{prandtl}            & Re$\sqrt{f/32}(k/r)$ & $f^{-1/2}$+2log($r/k$) & $c_{1} \left(c_{2} \left(c_{1}^{x} + c_{3}\right) + c_{4}^{x}\right)$ \\
She et~al.$^{c,d}$~\cite{she2012multi} & Re$\sqrt{f/32}(k/r)$ & $(1/\kappa)$ln(Re$\sqrt{f/32}$)+B-1/$(2\sqrt{f/32})$ & $\left(- c_{1} + \log{\left (c_{1}^{x} c_{2}^{2} + c_{3} + x \right )}\right) e^{c_{4}}$ \\
\end{tabular}
$^a C_s=3\times10^{-5}$ \\
$^b C_s=1\times10^{-8}$ and $\alpha=1/3+\eta/2$, with $\eta=0.02$ \\
$^c$ The independent variable defined by Prandtl and used by She et~al. is denoted k$_s^+$ in Figs.~\ref{fig:scalings}(d) and~\ref{fig:scalings}(e). \\
$^d$ Following She et al., this function is denoted $\Delta$U$^+_{\rm{avg}}$ in Fig.~\ref{fig:scalings}(e). B is a Reynolds-dependent correction applied for Re$\sqrt{f/32}<5000$ (see definition in~\cite{she2012multi}).
\end{ruledtabular}
\end{table*}

Let us first formalize the problem. The data collapse hypothesis posits that, under the appropriate transformation, all the observed data $D$ follow a single law $\fb = \hb(x)$ \footnote{In general, $x$ could be a vector of independent variables instead of just one variable as in Nikuradse's data.}. To identify the most plausible candidate for $\hb$, we start by considering the probability $p(h_i|D)$ that an expression $h_i$ is the correct one given the data, which can be written as \cite{robot}
\begin{equation}
 p(h_i|D) = \frac{\exp \left[-\dl(h_i)\right]}{Z} \; .
\end{equation}
Here, $Z=\sum_i \exp \left[-\dl(h_i) \right]$ is the partition function, and $\dl(h_i) = -\log p(h_i, D)$ is the description length of the model \cite{grunwald07,robot}. The description length plays the role of a free energy in a physical system, and the model with minimum description length is the most plausible one. The description length cannot, in general, be calculated exactly; however, it can be approximated as $\dl(h_i) = B(h_i)/2 - \log p(h_i)$, where $B(h_i)$ is the Bayesian information criterion of expression $h_i$ \cite{schwarz78,ando10} and the prior $p(h_i)$ is the probability assigned to $h_i$ before any data are observed. This prior acts as an expression regularizer and requires certain hypotheses; following previous work \cite{robot} we use the maximum entropy $p(h_i)$ that is consistent with empirically observed frequencies of each operation \cite{SM}.

We explore the space of possible mathematical expressions using the Metropolis algorithm, by means of what has been called a Bayesian machine scientist \cite{robot,SM}. The Bayesian machine scientist draws upon concepts developed in symbolic regression\cite{schmidt09}, exponential random graphs \cite{caimo11} and Bayesian network sampling \cite{horvat15,fischer15,robot}; it is guaranteed to asymptotically sample from the stationary distribution $p(h_i|D)$ and is consistent, that is, given enough data it will assign the highest plausibility (the shortest description length) to the correct model with probability approaching one. Among all models explored by the machine scientist using the Metropolis algorithm, we select the model with the minimum description length as the most plausible one \footnote{In practice, all sampled expressions describe the data similarly well, and none of the results below depend on which expressions we choose.}. Specifically, we let the machine scientist sample expressions for each of the collapses $(\fb, x)$ proposed for Nikuradse's data (Table~\ref{tab:scalings}).
From these samplings, we obtain the most plausible expression for each data collapse (Fig.~\ref{fig:scalings}(f)-(j)).

Despite fitting the collapsed data tightly, most of the functions do not reproduce the features of the original data when unscaled, particularly in the low and intermediate Reynolds regimes (Fig.~\ref{fig:scalings}(k)-(o)). This is important because, as mentioned earlier, collapse and scaling theories are usually evaluated by how close the data appear to be in the collapse. Our results for the Nikuradse data show how this can be misleading; some of the data collapses that have been proposed are effectively a \textit{zoom-out}, in which the curves for each roughness are stretched until the separation in the vertical axis is no longer visible. This can be achieved by multiplying both the scaling variable and the scaling function by a quantity that spans a broad range, typically a power of the Reynolds number. As we show here, a function fitting such a stretched curve does not necessarily recover the correct $f=h(\re,r/k)$ dependency when unscaled.

\begin{figure} 
  \centering
    \includegraphics[width=\columnwidth]{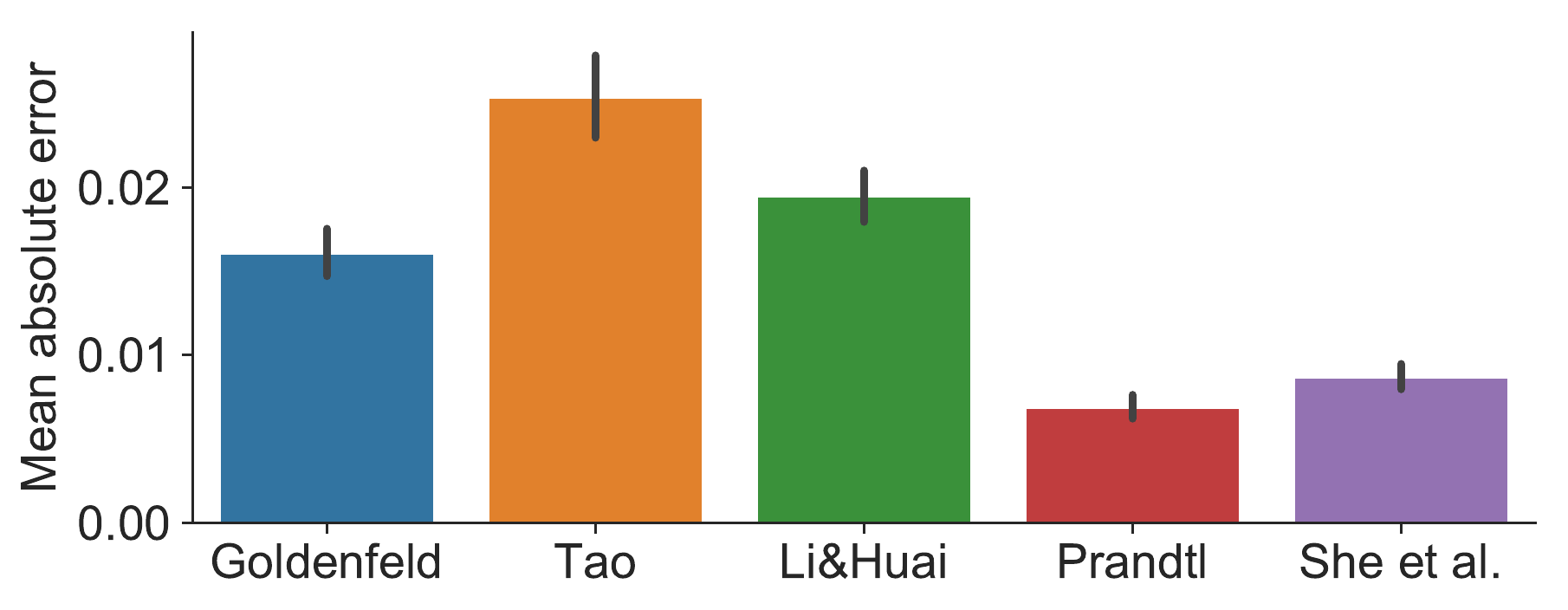}
  \vspace{-6mm}
  \caption{Mean absolute error (MAE) of the most plausible model for each data collapse, calculated on the original unscaled variables of the Nikuradse dataset (Fig.~\ref{fig:scalings}(k)-(o)). The error bars in indicate the 95\% confidence interval for the mean.}
\label{fig:errors}
\end{figure}

In Fig.~\ref{fig:errors}, we show the mean absolute error (MAE) for each of the scalings when the original variables are recovered. Unlike the deviations that one measures in the collapsed data, we argue that this quantity is a comparable and reliable measure of the true goodness of the collapse.
From this, we conclude that only the data collapses proposed by Prandtl~\cite{prandtl} and by She et al.~\cite{she2012multi} are accurate representations of the Nikuradse dataset in all regions. The remaining data collapses are only good representations of some regimes.

\begin{figure}
\centering
\includegraphics[width=\columnwidth]{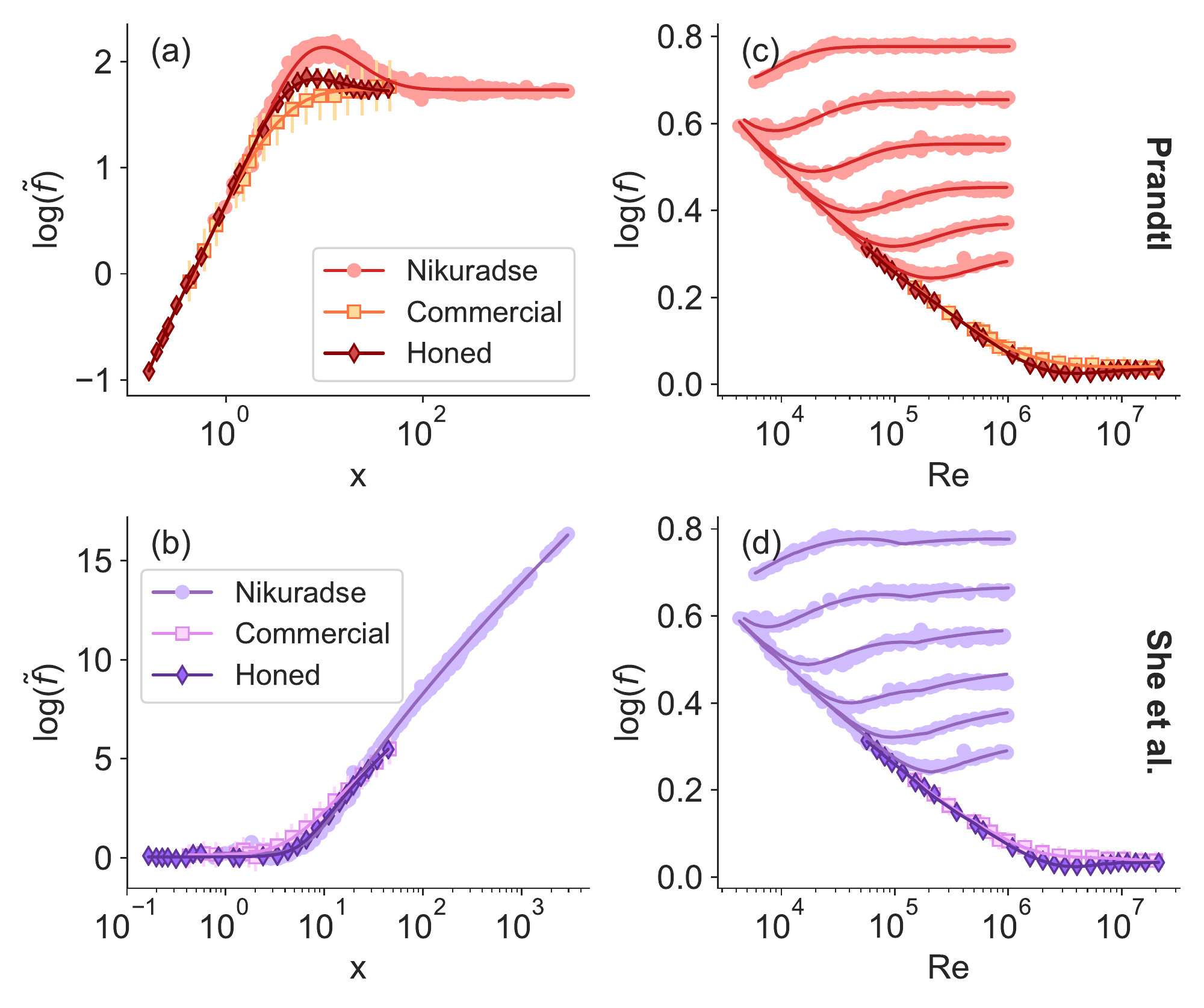}
\vspace{-8mm}
\caption{
(a)-(b) Models for the Nikuradse and Princeton (honed pipe with $k/r=8716$ and commercial pipe  with $k/r=8065$) datasets collapsed according to the approaches proposed by Prandtl~\cite{prandtl} (top) and She et al.~\cite{she2012multi} (bottom). The solid lines represent the most plausible closed-form mathematical model identified by the Bayesian machine scientist for all datasets simultaneously.
(c)-(d) Unscaled models.
}
\label{fig:princescaled}
\end{figure}

The collapses proposed by Prandtl~\cite{prandtl} and by She et al.~\cite{she2012multi} both use as their scaling variable the roughness Reynolds number $k_s^+$, and the scaling function depends, again on both cases, on the inverse square root of $f$. She et al. argued that the spread in the Prandtl-scaled data about $\mathrm{k}_s^+\sim1$ is too large for it to be considered a good collapse. Therefore, they introduced an extra parameter $p$, which encodes the behaviour at the transitionally rough regime, about $k_s^+\sim5$ in the scaled data (see Fig. 5 in \cite{she2012multi}). Effectively, $p$ is an exponent that tunes the sharpness of the transition between the different turbulent regimes with the roughness. Moreover, an ad-hoc correction was added to improve the collapse (the so-called $B$ term). This correction applies only to $\mathrm{Re}_\tau\equiv\mathrm{Re}\sqrt{f/32}<5000$ and thus makes the scaling function defined piecewise, introducing a number of extra parameters. Despite these refinements, our approach suggests that the Prandtl collapse still provides smoother and more accurate unscaled curves.

To further understand the validity of the original scaling by Prandtl and the need for the corrections introduced by She et al., we finally turn to the main criticism against Prandtl's scaling, namely that it does not hold for more recent empirical data from the Princeton experiments with honed \cite{honed} and commercial \cite{compipe} pipes. These experiments used smoother pipes than those used by Nikuradse ($k/r = 8716$ and $k/r = 8065$, respectively), and explored flows with $\mathrm{Re}$ up to $2\times10^7$. To investigate the applicability of Prandtl's collapse to these data, we check whether the machine scientist is able to generate a new set of expressions fitting both Nikuradse's and the Princeton data (Fig.~\ref{fig:princescaled}). Since the newer data seem to deviate from both Prandtl's and She et al.'s collapses (Fig.~4), we search for a single mathematical expression fitting the three datasets but with potentially different parameter values, accounting for different details arising, for example, from the geometry of the irregularities in each pipe.
This is possible thanks to the fact that the machine scientist samples mathematical expressions and, once an expression is selected, it can be fit separately to each dataset. The plausibility of the expression is then given by the total description length
\begin{equation}
 \dl(h_i) = \frac{1}{2} \sum_d B^d(h_i)-\log p(h_i) \;,
\end{equation}
where $B^d(h_i)$ is the Bayesian information criterion calculated on dataset $d$.

\begin{figure} 
  \centering
    \includegraphics[width=\columnwidth]{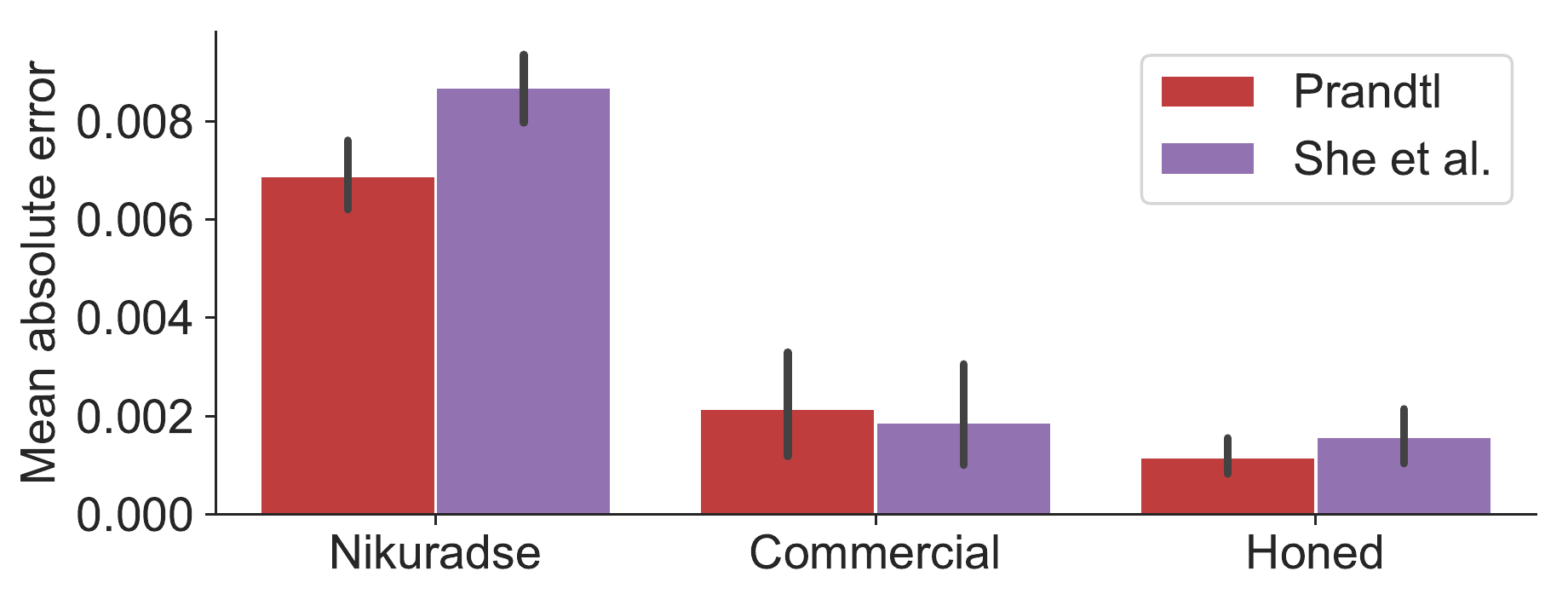}
  \vspace{-7mm}
  \caption{Mean absolute error (MAE) of the most plausible model for Prandtl's and for She et al.'s data collapses, calculated on the original unscaled variables of the Nikuradse dataset and the two Princeton datasets (right column in Fig.~\ref{fig:princescaled}). The error bars in indicate the 95\% confidence interval for the mean.}
\label{fig:pra-she_compare}
\end{figure}

%
%
The most plausible model fitting both the Nikuradse and the Princeton datasets using Prandtl's collapse is
\begin{equation}
\bar f\left(x\right) = \sinh{\left (c_{1} + \left(c_{2} + x\right) e^{- \left(c_{3} x^{2}\right)^{c_{4}}} \right )} \;, 
\label{eq:pramodel}
\end{equation}
whereas the most plausible model for She's collapse is
\begin{equation}
\bar f\left(x\right) = \left(c_{1} \log{\left (\left(c_{2} x^{c_{3}} + c_{4}\right)^{2} \right )}\right)^{c_{1}} \;.
\label{eq:shemodel}
\end{equation}
As above, the machine scientist uncovers, for each collapse, several models that are almost equally plausible. Therefore, the models in Eqs.~(\ref{eq:pramodel}) and (\ref{eq:shemodel}) should be taken as just an instance of the models describing each of the data collapses, and it is beyond our aims to discuss their putative physical meaning here. However, in both scaling laws we get two parameters whose value is shared (or very similar) among pipe types, whereas two others take pipe-specific values over a broader range (Tab.~\ref{tab:parames}).
Therefore, it seems plausible that two parameters are universal, whereas the other two may be dependent on properties of the material or the irregularities in it.
\begin{table}
\caption{}
\begin{ruledtabular}
\begin{tabular}{c|ccccc}
  \label{tab:parames}
scaling & Data set & $c_1$ & $c_2$ & $c_3$ & $c_4$ \\
\hline
\parbox[t]{2mm}{\multirow{3}{*}{\rotatebox[origin=c]{90}{Prandtl}}}
& Nikuradse  & 1.3 & -3.6 & 3.1             & 0.69 \\
& Honed      & 1.3 & -4.2 & 6.0             & 0.70 \\
& Commercial & 1.3 & -14  & 1.5$\times10^3$ & 0.62 \\
\hline
\parbox[t]{2mm}{\multirow{3}{*}{\rotatebox[origin=c]{90}{She}}} 
& Nikuradse  & 0.79 & 0.15 & 3.6 & 1.1 \\
& Honed      & 0.71 & 0.14 & 4.1 & 1.0 \\
& Commercial & 0.77 & 0.34 & 2.7 & 1.1 \\
\end{tabular}
\end{ruledtabular}
\end{table}
In any case, the addition of the Princeton experiments does not seem to justify the corrections introduced by She et al. Indeed, our approach suggests that the commercial and the honed pipes are equally well described by both data collapses (Fig.~\ref{fig:pra-she_compare}).

\bigskip

Machine learning tools are increasingly applied to shed light into physics problems \cite{zdeborova17}, from detecting phase transitions \cite{carrasquilla17,vannieuwenburg17} to approximating the wave function of many-body quantum systems \cite{carleo17}. Here, we have used a Bayesian machine scientist~\cite{robot}, which automatically uncovers closed-form mathematical equations from data, to compare data collapses for the Nikuradse dataset. The machine scientist enables us to sample a wealth of analytic expressions that describe each data collapse, using only probabilistic model-selection arguments and without introducing any heuristics or biases. We then use the expressions obtained by the Bayesian machine scientist to evaluate the goodness of the models on the unscaled data; as we show, the common practice of evaluating goodness of fit on the collapsed data leads to misleading conclusions.

In the Nikuradse dataset, our approach favors the original data collapse proposed by Prandtl over 80 years ago and, conversely, disfavors the data collapses proposed more recently ~\cite{goldenfeld2006roughness,tao2009critical,she2012multi,li2016united}. Our method suggests that these more recent approaches produce an apparent collapse of the data mainly by {\em compressing} the transition region between flow regimes; they only describe the experimental data in limiting regimes but not in these transition regions (although it is fair to note that their goal was, precisely, to describe some limiting behaviors and not all regimes). Our approach also shows that, contrary to what has been suggested, Prandtl's collapse is compatible with more recent experiments of turbulent friction, provided that some parameters in the scaling function are allowed to depend on the details of the pipe.

Although here we have focused on the Nikuradse dataset, we think that our approach can be applied to other systems on which data collapses are used to understand the underlying physics or mechanisms. Scalings and data collapses are, indeed, powerful tools; we argue that, combined with rigorous model selection and machine learning, they have the potential to become even more insightful.

\begin{acknowledgments}
We thank A. Arenas for pointing us towards the Nikuradse dataset. This project has received funding from the Spanish Ministerio de Economia y Competitividad (FIS2015-71563-ERC, FIS2016-78904-C3-P-1, and DPI2016-75791-C2-1-P) and from the Government of Catalonia (2017SGR-896).
\end{acknowledgments}


\begin{thebibliography}{32}%
\makeatletter
\providecommand \@ifxundefined [1]{%
 \@ifx{#1\undefined}
}%
\providecommand \@ifnum [1]{%
 \ifnum #1\expandafter \@firstoftwo
 \else \expandafter \@secondoftwo
 \fi
}%
\providecommand \@ifx [1]{%
 \ifx #1\expandafter \@firstoftwo
 \else \expandafter \@secondoftwo
 \fi
}%
\providecommand \natexlab [1]{#1}%
\providecommand \enquote  [1]{``#1''}%
\providecommand \bibnamefont  [1]{#1}%
\providecommand \bibfnamefont [1]{#1}%
\providecommand \citenamefont [1]{#1}%
\providecommand \href@noop [0]{\@secondoftwo}%
\providecommand \href [0]{\begingroup \@sanitize@url \@href}%
\providecommand \@href[1]{\@@startlink{#1}\@@href}%
\providecommand \@@href[1]{\endgroup#1\@@endlink}%
\providecommand \@sanitize@url [0]{\catcode `\\12\catcode `\$12\catcode
  `\&12\catcode `\#12\catcode `\^12\catcode `\_12\catcode `\%12\relax}%
\providecommand \@@startlink[1]{}%
\providecommand \@@endlink[0]{}%
\providecommand \url  [0]{\begingroup\@sanitize@url \@url }%
\providecommand \@url [1]{\endgroup\@href {#1}{\urlprefix }}%
\providecommand \urlprefix  [0]{URL }%
\providecommand \Eprint [0]{\href }%
\providecommand \doibase [0]{http://dx.doi.org/}%
\providecommand \selectlanguage [0]{\@gobble}%
\providecommand \bibinfo  [0]{\@secondoftwo}%
\providecommand \bibfield  [0]{\@secondoftwo}%
\providecommand \translation [1]{[#1]}%
\providecommand \BibitemOpen [0]{}%
\providecommand \bibitemStop [0]{}%
\providecommand \bibitemNoStop [0]{.\EOS\space}%
\providecommand \EOS [0]{\spacefactor3000\relax}%
\providecommand \BibitemShut  [1]{\csname bibitem#1\endcsname}%
\let\auto@bib@innerbib\@empty
\bibitem [{\citenamefont {Nikuradse}(1933)}]{nikuradse1933laws}%
  \BibitemOpen
  \bibfield  {author} {\bibinfo {author} {\bibfnamefont {J.}~\bibnamefont
  {Nikuradse}},\ }\href@noop {} {\bibfield  {journal} {\bibinfo  {journal}
  {Tech. Mem.}\ }\textbf {\bibinfo {volume} {1292}},\ \bibinfo {pages} {60}
  (\bibinfo {year} {1933})}\BibitemShut {NoStop}%
\bibitem [{\citenamefont {Prandtl}(1933)}]{prandtl}%
  \BibitemOpen
  \bibfield  {author} {\bibinfo {author} {\bibfnamefont {L.}~\bibnamefont
  {Prandtl}},\ }\href {https://ntrs.nasa.gov/search.jsp?R=19930094697} {\
  (\bibinfo {year} {1933})}\BibitemShut {NoStop}%
\bibitem [{Note1()}]{Note1}%
  \BibitemOpen
  \bibinfo {note} {In his original paper, Nikuradse already presented his data
  scaled according to Prandtl's collapse. However, this collapse is mentioned
  in~\cite {she2012multi}, where Bodenschatz attributes it to Prandtl,
  Nikuradse's supervisor.}\BibitemShut {Stop}%
\bibitem [{\citenamefont {Goldenfeld}(2006)}]{goldenfeld2006roughness}%
  \BibitemOpen
  \bibfield  {author} {\bibinfo {author} {\bibfnamefont {N.}~\bibnamefont
  {Goldenfeld}},\ }\href@noop {} {\bibfield  {journal} {\bibinfo  {journal}
  {Phys. Rev. Lett.}\ }\textbf {\bibinfo {volume} {96}},\ \bibinfo {pages}
  {044503} (\bibinfo {year} {2006})}\BibitemShut {NoStop}%
\bibitem [{\citenamefont {Tao}(2009)}]{tao2009critical}%
  \BibitemOpen
  \bibfield  {author} {\bibinfo {author} {\bibfnamefont {J.}~\bibnamefont
  {Tao}},\ }\href@noop {} {\bibfield  {journal} {\bibinfo  {journal} {Phys.
  Rev. Lett.}\ }\textbf {\bibinfo {volume} {103}},\ \bibinfo {pages} {264502}
  (\bibinfo {year} {2009})}\BibitemShut {NoStop}%
\bibitem [{\citenamefont {She}\ \emph {et~al.}(2012)\citenamefont {She},
  \citenamefont {Wu}, \citenamefont {Chen},\ and\ \citenamefont
  {Hussain}}]{she2012multi}%
  \BibitemOpen
  \bibfield  {author} {\bibinfo {author} {\bibfnamefont {Z.-S.}\ \bibnamefont
  {She}}, \bibinfo {author} {\bibfnamefont {Y.}~\bibnamefont {Wu}}, \bibinfo
  {author} {\bibfnamefont {X.}~\bibnamefont {Chen}}, \ and\ \bibinfo {author}
  {\bibfnamefont {F.}~\bibnamefont {Hussain}},\ }\href@noop {} {\bibfield
  {journal} {\bibinfo  {journal} {New J. Phys.}\ }\textbf {\bibinfo {volume}
  {14}},\ \bibinfo {pages} {093054} (\bibinfo {year} {2012})}\BibitemShut
  {NoStop}%
\bibitem [{\citenamefont {Li}\ and\ \citenamefont {Huai}(2016)}]{li2016united}%
  \BibitemOpen
  \bibfield  {author} {\bibinfo {author} {\bibfnamefont {S.}~\bibnamefont
  {Li}}\ and\ \bibinfo {author} {\bibfnamefont {W.}~\bibnamefont {Huai}},\
  }\href@noop {} {\bibfield  {journal} {\bibinfo  {journal} {PLoS ONE}\
  }\textbf {\bibinfo {volume} {11}},\ \bibinfo {pages} {e0154408} (\bibinfo
  {year} {2016})}\BibitemShut {NoStop}%
\bibitem [{\citenamefont {Stanley}(1999)}]{stanley99}%
  \BibitemOpen
  \bibfield  {author} {\bibinfo {author} {\bibfnamefont {H.~E.}\ \bibnamefont
  {Stanley}},\ }\href@noop {} {\bibfield  {journal} {\bibinfo  {journal} {Rev.
  Mod. Phys.}\ }\textbf {\bibinfo {volume} {71}},\ \bibinfo {pages} {S358}
  (\bibinfo {year} {1999})}\BibitemShut {NoStop}%
\bibitem [{\citenamefont {Barenblatt}(2003)}]{barenblatt03}%
  \BibitemOpen
  \bibfield  {author} {\bibinfo {author} {\bibfnamefont {G.~I.}\ \bibnamefont
  {Barenblatt}},\ }\href {\doibase 10.1017/CBO9780511814921} {\emph {\bibinfo
  {title} {Scaling}}},\ Cambridge Texts in Applied Mathematics\ (\bibinfo
  {publisher} {Cambridge University Press},\ \bibinfo {year}
  {2003})\BibitemShut {NoStop}%
\bibitem [{\citenamefont {Stanley}\ \emph {et~al.}(1996)\citenamefont
  {Stanley}, \citenamefont {Amaral}, \citenamefont {Buldyrev}, \citenamefont
  {Havlin}, \citenamefont {Leschhorn}, \citenamefont {Maass}, \citenamefont
  {Salinger},\ and\ \citenamefont {Stanley}}]{stanley96}%
  \BibitemOpen
  \bibfield  {author} {\bibinfo {author} {\bibfnamefont {M.~H.~R.}\
  \bibnamefont {Stanley}}, \bibinfo {author} {\bibfnamefont {L.~A.~N.}\
  \bibnamefont {Amaral}}, \bibinfo {author} {\bibfnamefont {S.~V.}\
  \bibnamefont {Buldyrev}}, \bibinfo {author} {\bibfnamefont {S.}~\bibnamefont
  {Havlin}}, \bibinfo {author} {\bibfnamefont {H.}~\bibnamefont {Leschhorn}},
  \bibinfo {author} {\bibfnamefont {P.}~\bibnamefont {Maass}}, \bibinfo
  {author} {\bibfnamefont {M.~A.}\ \bibnamefont {Salinger}}, \ and\ \bibinfo
  {author} {\bibfnamefont {H.~E.}\ \bibnamefont {Stanley}},\ }\href@noop {}
  {\bibfield  {journal} {\bibinfo  {journal} {Nature}\ }\textbf {\bibinfo
  {volume} {379}},\ \bibinfo {pages} {804} (\bibinfo {year}
  {1996})}\BibitemShut {NoStop}%
\bibitem [{\citenamefont {Lee}\ \emph {et~al.}(1998)\citenamefont {Lee},
  \citenamefont {Amaral}, \citenamefont {Meyer}, \citenamefont {Canning},\ and\
  \citenamefont {Stanley}}]{lee98}%
  \BibitemOpen
  \bibfield  {author} {\bibinfo {author} {\bibfnamefont {Y.}~\bibnamefont
  {Lee}}, \bibinfo {author} {\bibfnamefont {L.~A.~N.}\ \bibnamefont {Amaral}},
  \bibinfo {author} {\bibfnamefont {M.}~\bibnamefont {Meyer}}, \bibinfo
  {author} {\bibfnamefont {D.}~\bibnamefont {Canning}}, \ and\ \bibinfo
  {author} {\bibfnamefont {H.~E.}\ \bibnamefont {Stanley}},\ }\href@noop {}
  {\bibfield  {journal} {\bibinfo  {journal} {Phys. Rev. Lett.}\ }\textbf
  {\bibinfo {volume} {81}},\ \bibinfo {pages} {3275} (\bibinfo {year}
  {1998})}\BibitemShut {NoStop}%
\bibitem [{Note2()}]{Note2}%
  \BibitemOpen
  \bibinfo {note} {We refer to $\protect \mathaccentV {bar}016{h}$ as the
  {\protect \em scaling function} although, strictly speaking, this function is
  not always a generalized homogeneous function.}\BibitemShut {Stop}%
\bibitem [{\citenamefont {Bhattacharjee}\ and\ \citenamefont
  {Seno}(2001)}]{bhattacharjee01}%
  \BibitemOpen
  \bibfield  {author} {\bibinfo {author} {\bibfnamefont {S.~M.}\ \bibnamefont
  {Bhattacharjee}}\ and\ \bibinfo {author} {\bibfnamefont {F.}~\bibnamefont
  {Seno}},\ }\href@noop {} {\bibfield  {journal} {\bibinfo  {journal} {J. Phys.
  A}\ }\textbf {\bibinfo {volume} {34}},\ \bibinfo {pages} {6375} (\bibinfo
  {year} {2001})}\BibitemShut {NoStop}%
\bibitem [{\citenamefont {Evans}\ and\ \citenamefont
  {Rzhetsky}(2010)}]{evans10}%
  \BibitemOpen
  \bibfield  {author} {\bibinfo {author} {\bibfnamefont {J.}~\bibnamefont
  {Evans}}\ and\ \bibinfo {author} {\bibfnamefont {A.}~\bibnamefont
  {Rzhetsky}},\ }\href@noop {} {\bibfield  {journal} {\bibinfo  {journal}
  {Science}\ }\textbf {\bibinfo {volume} {329}},\ \bibinfo {pages} {399}
  (\bibinfo {year} {2010})}\BibitemShut {NoStop}%
\bibitem [{\citenamefont {Guimer\`a}\ \emph {et~al.}(2020)\citenamefont
  {Guimer\`a}, \citenamefont {Reichardt}, \citenamefont {Aguilar-Mogas},
  \citenamefont {Massucci}, \citenamefont {Miranda}, \citenamefont
  {Pallar\`es},\ and\ \citenamefont {Sales-Pardo}}]{robot}%
  \BibitemOpen
  \bibfield  {author} {\bibinfo {author} {\bibfnamefont {R.}~\bibnamefont
  {Guimer\`a}}, \bibinfo {author} {\bibfnamefont {I.}~\bibnamefont
  {Reichardt}}, \bibinfo {author} {\bibfnamefont {A.}~\bibnamefont
  {Aguilar-Mogas}}, \bibinfo {author} {\bibfnamefont {F.~A.}\ \bibnamefont
  {Massucci}}, \bibinfo {author} {\bibfnamefont {M.}~\bibnamefont {Miranda}},
  \bibinfo {author} {\bibfnamefont {J.}~\bibnamefont {Pallar\`es}}, \ and\
  \bibinfo {author} {\bibfnamefont {M.}~\bibnamefont {Sales-Pardo}},\
  }\href@noop {} {\bibfield  {journal} {\bibinfo  {journal} {Sci. Adv.}\
  }\textbf {\bibinfo {volume} {6}},\ \bibinfo {pages} {eaav6971} (\bibinfo
  {year} {2020})}\BibitemShut {NoStop}%
\bibitem [{Note3()}]{Note3}%
  \BibitemOpen
  \bibinfo {note} {In general, $x$ could be a vector of independent variables
  instead of just one variable as in Nikuradse's data.}\BibitemShut {Stop}%
\bibitem [{\citenamefont {Gr\"{u}nwald}(2007)}]{grunwald07}%
  \BibitemOpen
  \bibfield  {author} {\bibinfo {author} {\bibfnamefont {P.~D.}\ \bibnamefont
  {Gr\"{u}nwald}},\ }\href@noop {} {\emph {\bibinfo {title} {The Minimum
  Description Length Principle}}}\ (\bibinfo  {publisher} {The MIT Press},\
  \bibinfo {address} {Cambridge, Massachusetts},\ \bibinfo {year}
  {2007})\BibitemShut {NoStop}%
\bibitem [{\citenamefont {Schwarz}(1978)}]{schwarz78}%
  \BibitemOpen
  \bibfield  {author} {\bibinfo {author} {\bibfnamefont {G.}~\bibnamefont
  {Schwarz}},\ }\href@noop {} {\bibfield  {journal} {\bibinfo  {journal} {Ann.
  Stat.}\ }\textbf {\bibinfo {volume} {6}},\ \bibinfo {pages} {461} (\bibinfo
  {year} {1978})}\BibitemShut {NoStop}%
\bibitem [{\citenamefont {Ando}(2010)}]{ando10}%
  \BibitemOpen
  \bibfield  {author} {\bibinfo {author} {\bibfnamefont {T.}~\bibnamefont
  {Ando}},\ }\href@noop {} {\emph {\bibinfo {title} {Bayesian model selection
  and statistical modeling}}}\ (\bibinfo  {publisher} {CRC Press},\ \bibinfo
  {year} {2010})\BibitemShut {NoStop}%
\bibitem [{SM()}]{SM}%
  \BibitemOpen
  \href@noop {} {}\bibinfo {note} {See Supplemental Material, which includes
  Ref.~\cite{earl05}.}\BibitemShut {Stop}%
\bibitem [{\citenamefont {Schmidt}\ and\ \citenamefont
  {Lipson}(2009)}]{schmidt09}%
  \BibitemOpen
  \bibfield  {author} {\bibinfo {author} {\bibfnamefont {M.}~\bibnamefont
  {Schmidt}}\ and\ \bibinfo {author} {\bibfnamefont {H.}~\bibnamefont
  {Lipson}},\ }\href {\doibase 10.1126/science.1165893} {\bibfield  {journal}
  {\bibinfo  {journal} {Science}\ }\textbf {\bibinfo {volume} {324}},\ \bibinfo
  {pages} {81} (\bibinfo {year} {2009})}\BibitemShut {NoStop}%
\bibitem [{\citenamefont {Caimo}\ and\ \citenamefont {Friel}(2011)}]{caimo11}%
  \BibitemOpen
  \bibfield  {author} {\bibinfo {author} {\bibfnamefont {A.}~\bibnamefont
  {Caimo}}\ and\ \bibinfo {author} {\bibfnamefont {N.}~\bibnamefont {Friel}},\
  }\href@noop {} {\bibfield  {journal} {\bibinfo  {journal} {Soc. Netw.}\
  }\textbf {\bibinfo {volume} {33}},\ \bibinfo {pages} {41} (\bibinfo {year}
  {2011})}\BibitemShut {NoStop}%
\bibitem [{\citenamefont {Horv\'at}\ \emph {et~al.}(2015)\citenamefont
  {Horv\'at}, \citenamefont {Czabarka},\ and\ \citenamefont
  {Toroczkai}}]{horvat15}%
  \BibitemOpen
  \bibfield  {author} {\bibinfo {author} {\bibfnamefont {S.}~\bibnamefont
  {Horv\'at}}, \bibinfo {author} {\bibfnamefont {E.}~\bibnamefont {Czabarka}},
  \ and\ \bibinfo {author} {\bibfnamefont {Z.}~\bibnamefont {Toroczkai}},\
  }\href {\doibase 10.1103/PhysRevLett.114.158701} {\bibfield  {journal}
  {\bibinfo  {journal} {Phys. Rev. Lett.}\ }\textbf {\bibinfo {volume} {114}},\
  \bibinfo {pages} {158701} (\bibinfo {year} {2015})}\BibitemShut {NoStop}%
\bibitem [{\citenamefont {Fischer}\ \emph {et~al.}(2015)\citenamefont
  {Fischer}, \citenamefont {Leit\~ao}, \citenamefont {Peixoto},\ and\
  \citenamefont {Altmann}}]{fischer15}%
  \BibitemOpen
  \bibfield  {author} {\bibinfo {author} {\bibfnamefont {R.}~\bibnamefont
  {Fischer}}, \bibinfo {author} {\bibfnamefont {J.~C.}\ \bibnamefont
  {Leit\~ao}}, \bibinfo {author} {\bibfnamefont {T.~P.}\ \bibnamefont
  {Peixoto}}, \ and\ \bibinfo {author} {\bibfnamefont {E.~G.}\ \bibnamefont
  {Altmann}},\ }\href {\doibase 10.1103/PhysRevLett.115.188701} {\bibfield
  {journal} {\bibinfo  {journal} {Phys. Rev. Lett.}\ }\textbf {\bibinfo
  {volume} {115}},\ \bibinfo {pages} {188701} (\bibinfo {year}
  {2015})}\BibitemShut {NoStop}%
\bibitem [{Note4()}]{Note4}%
  \BibitemOpen
  \bibinfo {note} {In practice, all sampled expressions describe the data
  similarly well, and none of the results below depend on which expressions we
  choose.}\BibitemShut {Stop}%
\bibitem [{\citenamefont {Shockling}\ \emph {et~al.}(2006)\citenamefont
  {Shockling}, \citenamefont {Allen},\ and\ \citenamefont {Smits}}]{honed}%
  \BibitemOpen
  \bibfield  {author} {\bibinfo {author} {\bibfnamefont {M.~A.}\ \bibnamefont
  {Shockling}}, \bibinfo {author} {\bibfnamefont {J.~J.}\ \bibnamefont
  {Allen}}, \ and\ \bibinfo {author} {\bibfnamefont {A.~J.}\ \bibnamefont
  {Smits}},\ }\href {\doibase 10.1017/S0022112006001467} {\bibfield  {journal}
  {\bibinfo  {journal} {J. Fluid Mech.}\ }\textbf {\bibinfo {volume} {564}},\
  \bibinfo {pages} {267–285} (\bibinfo {year} {2006})}\BibitemShut {NoStop}%
\bibitem [{\citenamefont {Langelandsvik}\ \emph {et~al.}(2008)\citenamefont
  {Langelandsvik}, \citenamefont {Kunkel},\ and\ \citenamefont
  {Smits}}]{compipe}%
  \BibitemOpen
  \bibfield  {author} {\bibinfo {author} {\bibfnamefont {L.~I.}\ \bibnamefont
  {Langelandsvik}}, \bibinfo {author} {\bibfnamefont {G.~J.}\ \bibnamefont
  {Kunkel}}, \ and\ \bibinfo {author} {\bibfnamefont {A.~J.}\ \bibnamefont
  {Smits}},\ }\href {\doibase 10.1017/S0022112007009305} {\bibfield  {journal}
  {\bibinfo  {journal} {J. Fluid Mech.}\ }\textbf {\bibinfo {volume} {595}},\
  \bibinfo {pages} {323–339} (\bibinfo {year} {2008})}\BibitemShut {NoStop}%
\bibitem [{\citenamefont {Zdeborov{\'{a}}}(2017)}]{zdeborova17}%
  \BibitemOpen
  \bibfield  {author} {\bibinfo {author} {\bibfnamefont {L.}~\bibnamefont
  {Zdeborov{\'{a}}}},\ }\href {\doibase 10.1038/nphys4053} {\bibfield
  {journal} {\bibinfo  {journal} {Nat. Phys.}\ }\textbf {\bibinfo {volume}
  {13}},\ \bibinfo {pages} {420} (\bibinfo {year} {2017})}\BibitemShut
  {NoStop}%
\bibitem [{\citenamefont {Carrasquilla}\ and\ \citenamefont
  {Melko}(2017)}]{carrasquilla17}%
  \BibitemOpen
  \bibfield  {author} {\bibinfo {author} {\bibfnamefont {J.}~\bibnamefont
  {Carrasquilla}}\ and\ \bibinfo {author} {\bibfnamefont {R.~G.}\ \bibnamefont
  {Melko}},\ }\href@noop {} {\bibfield  {journal} {\bibinfo  {journal} {Nat.
  Phys.}\ }\textbf {\bibinfo {volume} {13}},\ \bibinfo {pages} {431} (\bibinfo
  {year} {2017})}\BibitemShut {NoStop}%
\bibitem [{\citenamefont {van Nieuwenburg}\ \emph {et~al.}(2017)\citenamefont
  {van Nieuwenburg}, \citenamefont {Liu},\ and\ \citenamefont
  {Huber}}]{vannieuwenburg17}%
  \BibitemOpen
  \bibfield  {author} {\bibinfo {author} {\bibfnamefont {E.~P.~L.}\
  \bibnamefont {van Nieuwenburg}}, \bibinfo {author} {\bibfnamefont {Y.-H.}\
  \bibnamefont {Liu}}, \ and\ \bibinfo {author} {\bibfnamefont {S.~D.}\
  \bibnamefont {Huber}},\ }\href@noop {} {\bibfield  {journal} {\bibinfo
  {journal} {Nat. Phys.}\ }\textbf {\bibinfo {volume} {13}},\ \bibinfo {pages}
  {435} (\bibinfo {year} {2017})}\BibitemShut {NoStop}%
\bibitem [{\citenamefont {Carleo}\ and\ \citenamefont
  {Troyer}(2017)}]{carleo17}%
  \BibitemOpen
  \bibfield  {author} {\bibinfo {author} {\bibfnamefont {G.}~\bibnamefont
  {Carleo}}\ and\ \bibinfo {author} {\bibfnamefont {M.}~\bibnamefont
  {Troyer}},\ }\href@noop {} {\bibfield  {journal} {\bibinfo  {journal}
  {Science}\ }\textbf {\bibinfo {volume} {355}},\ \bibinfo {pages} {602}
  (\bibinfo {year} {2017})}\BibitemShut {NoStop}%
\bibitem [{\citenamefont {Earl}\ and\ \citenamefont {Deem}(2005)}]{earl05}%
  \BibitemOpen
  \bibfield  {author} {\bibinfo {author} {\bibfnamefont {D.~J.}\ \bibnamefont
  {Earl}}\ and\ \bibinfo {author} {\bibfnamefont {M.~W.}\ \bibnamefont
  {Deem}},\ }\href@noop {} {\bibfield  {journal} {\bibinfo  {journal} {Phys.
  Chem. Chem. Phys.}\ }\textbf {\bibinfo {volume} {7}},\ \bibinfo {pages}
  {3910} (\bibinfo {year} {2005})}\BibitemShut {NoStop}%
\end{thebibliography}

%

\end{document}